\documentclass[prb,aps,reprint]{revtex4-1}

\usepackage{graphicx}
\usepackage{amssymb}
\usepackage{amsmath}

\begin{document}
\title{Strong spin-orbit coupling and magnetism in (111) (La$_{0.3}$Sr$_{0.7}$)(Al$_{0.65}$Ta$_{0.35})$/SrTiO$_3$}
\author{V. V. Bal$^1$, Z. Huang$^{2,3}$, K. Han$^{2,3}$, Ariando$^{2,3,4}$, T. Venkatesan$^{2,3,4,5,6}$ and V. Chandrasekhar$^{*}$}
\affiliation{$^1$Department of Physics, Northwestern University, Evanston, IL 60208, USA,\\ 
 $^2$ NUSNNI-Nanocore, National University of Singapore 117411, Singapore\\
 $3$ Department of Physics, National University of Singapore 117551, Singapore\\
 $4$ NUS Graduate School for Integrative Sciences and Engineering, National University of Singapore 117456, Singapore\\
 $5$ Department of Electrical and Computer Engineering, National University of Singapore 117576, Singapore\\
 $6$ Department of Material Science and Engineering, National University of Singapore 117575, Singapore}
 \date{\today}             
 \begin{abstract}
Two dimensional conducting interfaces in SrTiO$_3$ based heterostructures display a variety of coexisting and competing physical phenomena, which can be tuned by the application of a gate voltage. 
(111) oriented heterostructures have recently gained attention due to the possibility of finding exotic physics in these systems owing to their hexagonal surface crystal symmetry.  In this work, we use magnetoresistance to study the evolution of spin-orbit interaction and magnetism in (111) oriented (La$_{0.3}$Sr$_{0.7}$)(Al$_{0.65}$Ta$_{0.35}$)/SrTiO$_3$.  At more positive values of the gate voltage, which correspond to high carrier densities, we find that transport is multiband, and dominated by high mobility carriers with a tendency towards weak localization.  At more negative gate voltages, the carrier density is reduced, the high mobility bands are depopulated, and weak antilocalization effects begin to dominate, indicating that spin-orbit interaction becomes stronger.  At millikelvin temperatures, at gate voltages corresponding to the strong spin-orbit regime, we observe hysteresis in magnetoresistance, indicative of ferromagnetism in the system.  Our results suggest that in the (111) (La$_{0.3}$Sr$_{0.7}$)(Al$_{0.65}$Ta$_{0.35}$)/SrTiO$_3$ system, low mobility carriers which experience strong spin-orbit interactions participate in creating magnetic order in the system.
\end{abstract} 
\maketitle

The two dimensional carrier gas (2DCG) in SrTiO$_3$ (STO) based heterostructures has been a source of great research interest since its discovery in 2004,\cite{Ohtomo} owing to the variety of tunable physical phenomena it displays.\cite{Brinkman, Dikin, Bi,Thiel} 
A multitude of experimental probes have revealed magnetic behavior in these systems.  In particular, magnetoresistance (MR) measurements have shown signatures of the Kondo effect,\cite{Brinkman, Goldhaber} anisotropic MR \cite{Fete, Annadi2, Dagansixfold} and the anomalous Hall effect without hysteresis,\cite{YLee} as well as hysteretic MR,\cite{Brinkman, Dikin, Mehta, moetkalf, panagopoulos} with varying temperature ($T$) and gate voltage ($V_g$) dependencies.  Although there is no consensus yet, theories of magnetism rely on an understanding of the band structure to determine interactions between local moments and/or itinerant electrons.  One of the important parameters affecting the band structure is the spin-orbit interaction (SOI), which has been shown to be strong in STO based 2DCGs. This SOI, which is tunable by the application of $V_g$,\cite{CavigliaSOC} can also play a more direct role in determining the ground state of the system, by giving rise to various spin textures,\cite{Banerjee} which are very interesting from the point of view of applications as well as fundamental physics.\cite{Fert}

So far, the band structure has been widely studied in the case of (001) oriented STO based systems.  The degeneracy of the Ti 3$d$ $t_{2g}$ orbitals in (001) STO is broken near the interface, with the in-plane $d_{xy}$ orbitals of lowest energy, while the $d_{yz}$ and $d_{zx}$ orbitals are higher in energy.\cite{Syroold}  The electrons initially occupy the $d_{xy}$ bands, but as the carrier density is increased by increasing $V_g$, the $d_{yz}$ and $d_{zx}$ bands start to become occupied.\cite{Joshua}  In (001) LAO/STO devices, the SOI increases as $V_g$ is increased, as demonstrated by MR studies in which the low field MR goes from negative to positive as $V_g$ is increased.\cite{CavigliaSOC} The band structure in the case of (111) oriented STO based systems is very different, and shows a hexagonal symmetry of the Ti 3$d$ $t_{2g}$ bands.\cite{Syro, Walker, Pickett, Xiao}  According to angle-resolved photoemission (ARPES) studies on vacuum cleaved (111) STO,\cite{Walker} these bands are almost degenerate at the $\Gamma$ point. First principle studies on (111) oriented perovskite heterostructures show that the trigonal crystal field and SOI can split the  degeneracy of the Ti $3d$ $t_{2g}$ manifold, and that the orbital ordering is sensitively dependent on strain.\cite{Pickett,Xiao}  How this complex band structure affects various properties is just beginning to be explored.\cite{Annadi,Herranz}  Recent transport experiments on (111) LaAlO$_3$/STO (LAO/STO) have shown surprising anisotropies related to the crystalline axes, dependent on an applied $V_g$,\cite{Davis, Davis2, Dagansixfold,Davis3,Davis4,Davis5} as well as a non-monotonic dependence of phase coherence time $\tau_\phi$ and spin-orbit scattering time $\tau_{so}$ on $V_g$.\cite{Rout}  Here we report studies of the MR of LSAT deposited epitaxially on (111) STO.   LSAT has a 1\% lattice mismatch with STO, as opposed to a 3\% mismatch in the case of LAO/STO,\cite{Huang} which leads to a comparatively lower strain.  As with the (001) LAO/STO heterostructures, we find that the SOI can be tuned by $V_g$. Unlike the (001) LAO/STO heterostructures,\cite{CavigliaSOC} the SOI \emph{increases} as $V_g$ is tuned to more negative values.  Along with this increase in SOI, at millikelvin temperatures, we see an onset of hysteresis in the MR, a signature of magnetic order.  This correlation suggests that SOI may play an important role in the magnetism that has been observed in STO-based heterostructures. 

\begin{figure}[b!]
      \centering
      \includegraphics[width = .31\textwidth]{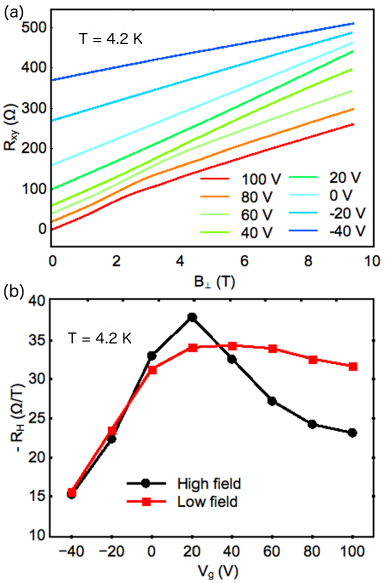}
      \caption{\textbf{a,} Hall resistance for various values of $V_g$ at $T$ = 4.2 K. Data have been antisymmetrized to remove a small contribution of longitudinal resistance due to a slight mismatch of the position of Hall probes along the length of the Hall bar. Data are also displaced along the vertical axis for clarity. \textbf{b,} Low field (red squares) and high field (black dots) Hall slopes for data taken at 4.2 K. The low field and high field values correspond to values of the derivative of the Hall resistance with respect to $B_\perp$, at $B_\perp$ = 0 and 9 T, respectively.}
      \label{fig1}
\end{figure}

We measured the MR of four Hall bars fabricated using photolithography and Ar ion milling on a single chip, with two of them aligned along the [1$\bar{1}$0] surface crystal direction, and the other two along the [$\bar{1}\bar{1}$2] direction.  The length of the Hall bars was 600 $\mu$m and the width was 100 $\mu$m.  
The sample had 12 monolayers of LSAT deposited on (111) oriented STO using pulsed laser deposition, at a growth partial oxygen pressure of 10$^{-4}$ Torr.\cite{Huang,Huang2}  No post growth annealing step was performed.  Ti/Au was deposited on contact pads of the Hall bars, and Al wirebonds were made to allow for a 4-probe measurement of transverse and longitudinal resistance.  The sample was attached to a copper puck using silver paint,
which allowed for the application of $V_g$.  Measurements in field perpendicular to the sample plane were performed in an Oxford Kelvinox 300 dilution refrigerator, while measurements in field parallel to the sample plane were carried out in an Oxford Kelvinox MX100 refrigerator.  Standard lock-in measurement techniques were used, with an ac frequency of 3 Hz, and an ac current $\sim$ 100 nA.  The data for all the Hall bars were qualitatively similar \cite{Bal} and the trends observed were reproducible over multiple cool-downs:  unlike the (111) LAO/STO devices \cite{Davis} we did not observe systematic differences between Hall bars aligned along the two crystal directions.  Consequently, here we show detailed data for only one Hall bar aligned along the [$\bar{1}\bar{1}$2] direction.  In order to maintain a uniform protocol, $V_g$ was swept multiple times between 100 V and -40 V at 4.2 K to make sure the sample was in the reversible electrostatic doping regime,\cite{Biscaras} after which the MR was measured at various values of $V_g$, changing $V_g$ in steps of 20 V, going always from 100 V to -40 V.

\begin{figure}[b!]
      \centering
      \includegraphics[width = .31\textwidth]{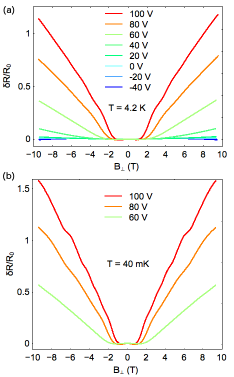}
      \caption{\textbf{a,} Differential MR at $T$ = 4.2 K for various values of $V_g$. \textbf{b,} Differential MR for $V_g$ = 100, 80, and 60 V, at $T$ = 40 mK.}  
     \label{fig2}
\end{figure}
Figure \ref{fig1}(a) shows the transverse Hall resistance $R_{xy}$ at $T$ = 4.2 K for various values of $V_g$. 
The sign of the Hall resistance is electron-like for all values of $V_g$ in this study.  At higher values of $V_g$, the Hall resistance is nonlinear:  a knee exists in the Hall response for $V_g$  = 100 V at $B_\perp$ $\sim$ 2.5 T, and moves to higher fields as $V_g$ is decreased. The Hall response appears to be linear for $V_g$ $\lesssim$ 20 V. This transition from a nonlinear to linear Hall effect as $V_g$ is decreased is similar to what is observed in the case of (001) STO based systems,\cite{Joshua} and has been interpreted there as a transition from multi-carrier transport to single carrier transport.  Based on the expected band structure from simple tight-binding approximations as well as ARPES studies,\cite{Syro,Walker} one expects two electron bands of the $t_{2g}$ manifold to be degenerate with one effective mass, while the remaining electron band has a different effective mass.  Consequently, we find that it is reasonable to describe the Hall data in terms of a two-electron band model with two different densities $n_1$ and $n_2$, and two different mobilities $\mu_1$ and $\mu_2$, at least for the more positive values of $V_g$ (see Supplementary Information).  At low magnetic fields, the resulting Hall coefficient $R_H$ is a function of the mobilities and the densities of both bands, but in the limit of large magnetic fields, it is determined solely by the total electron density, $|R_H|\sim 1/(n_1 + n_2)$.  Consequently one should be able to qualitatively determine the dependence of the charge density by examining the behavior of the high-field Hall coefficient as a function of $V_g$. 

Figure \ref{fig1}(b) shows the low and high field values of $R_H = dR_{xy}/dB_\perp$ as a function of $V_g$.  For $V_g>20$ V, the high-field $|R_H|$ decreases with increasing $V_g$.  Since the density of electrons increases with increasing $V_g$, this dependence is expected.  For $V_g<20$ V, however, the high-field $|R_H|$ decreases as $V_g$ is reduced.  This dependence is inconsistent with the presence of only electron-like bands discussed above, even if the mobilities of the electron bands are reduced with decreasing $V_g$.  In the case of (111) LAO/STO devices, a similar dependence was observed \cite{Davis}, where it was ascribed to the presence of holes in additions to electrons at the interface.  We believe a similar situation exists in these (111) LSAT/STO devices. Low mobility hole-like carriers are important at lower values of $V_g$, and may be present even at higher values of $V_g$. The low field $R_H$ has a qualitatively similar (albeit weaker) dependence on $V_g$, and low-field and high-field $R_H$ begin to coincide for $V_g<20$ V, reflecting the fact that the Hall resistance is linear over the entire range of field at these gate voltages.       

Evidence for multiband transport is also found in the longitudinal MR data shown in Figs. \ref{fig2}(a) and (b).  For $V_g$=80 and 100 V, the MR is large and positive, and shows evidence of Shubnikov-de Haas (SdH) oscillations, which are more pronounced at lower temperatures.  Such oscillations are expected when $\omega_c\tau$ $>$ 1, where $\omega_c$ = $eB_\perp/m^*$ is the cyclotron frequency, $m^*$ is the effective mass and $\tau$ is the scattering time.   The fact that SdH oscillations are seen starting at relatively low fields ($B_\perp \sim 1.5$ T) indicates that the mobility of the carriers for large positive $V_g$ is large.  Analysis of the oscillations between $\sim$2 and $\sim$9 T shows two dominant periods (see Supplementary Information), consistent with the presence of two kinds of high mobility carriers present at high values of $V_g$.  Near zero field, one observes a negative MR, consistent with weak localization effects, discussed in more detail later.  We note that the high field MR does not saturate, as would be expected for a system with closed orbits and only electron-like carriers, but instead shows a quasilinear increase.\cite{Pippard}  In our sample, we believe the likely causes for the quasilinear MR are some degree of compensation,\cite{Pippard} as suggested by the Hall data, and the presence of sample inhomogeneities,\cite{Parish} which are unavoidable in these perovskite samples. Our present data does not allow us to distinguish between the two mechanisms. 

As $V_g$ is reduced, the overall magnitude of the MR decreases dramatically, and SdH oscillations disappear.  The absence of SdH oscillations is expected, as the resistivity increases and $\mu$ decreases rapidly with decreasing $V_g$ (see Supplementary Information).  Semiclassically, for a single band system the longitudinal resistances vanishes. The MR is also expected to be very small for low mobility carriers. We believe that the rapid decrease in the magnitude of the longitudinal MR below $V_g \sim 20$ V is associated with the concomitant depopulation of high mobility electron bands that were occupied at higher gate voltages.  Thus transport below $V_g \sim$ 20 V is due to low mobility electrons and holes. The remanent MR is small, and as we explain below, associated with quantum interference corrections to the resistance.  

\begin{figure*}[ht!]
      \begin{center}
      \includegraphics[width = 0.65\textwidth]{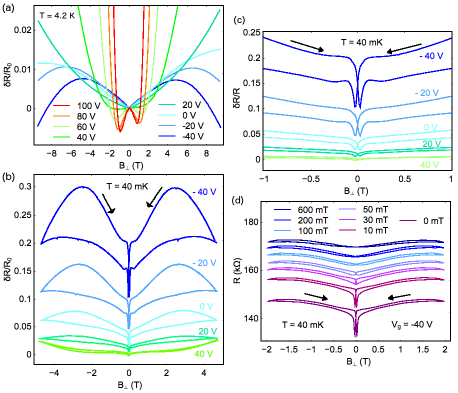}
      \caption{\textbf{a,} Differential MR at $T$ = 4.2 K for various value of $V_g$. \textbf{b,} Differential MR at $T$ = 40 mK, for $V_g$ $<$ 60 V. The data are shifted along the vertical axis for clarity. Arrows show the direction of magnetic field sweep, swept at the rate of 2.78 mT/s.  \textbf{c,} Same data as in Fig. 3b, zoomed in to lower field values. \textbf{d,} MR as a function of $B_\perp$ swept at the rate of 1.2 mT/s, measured with various values of in-plane field held constant for $V_g$ = -40 V, at $T$ = 40 mK.}
      \label{fig3}
     \end{center}
\end{figure*}

We now focus on the low field MR at 4.2 K, shown in Fig. \ref{fig3}a.  The longitudinal MR at low fields is negative for $V_g > 20$ V, and positive below this value.  In LAO/STO heterostructures, the low field MR has been associated with coherent backscattering, i.e., weak localization effects, and we believe that the origin of the low field MR here is the same.  In the presence of spin-dependent scattering, the weak localization contribution to the normalized differential resistance can be expressed as:\cite{Hikami}
\begin{equation}
\frac{\delta R}{R_0} = \frac{R(B) - R(0)}{R(0)} = -\frac{3}{2}f(\ell_2,B) + \frac{1}{2}f(\ell_1, B)
\label{eqn1}
\end{equation} 
where the first term is the triplet Cooperon contribution, and the second the singlet contribution.  Here $\ell_2^{-2} = \ell_{\phi}^{-2} + (4/3)\ell_{so}^{-2} + (2/3) \ell_s^{-2}$ and $\ell_1^{-2} = \ell_{\phi}^{-2} +  2\ell_s^{-2}$, $\ell_{\phi}$, $\ell_{so}$ and $\ell_s$ being the electron phase coherence length, the spin-orbit scattering length and the magnetic scattering length respectively.  In a quasi-two-dimensional system, the function $f$, written in terms of the characteristic fields $B_i = (h/2e)/(2 \pi \ell_i^2)$, is given by:\cite{Santhanam}
\begin{equation}
f(B_i, B) =  \frac{e^2 R_\square}{2 \pi^2 \hbar} \Big[ \Psi \big( \frac{1}{2} + \frac{B_i}{B} \big) - \mathrm{ln} \big(\frac{B_i}{B} \big) \Big].
\label{eqn2}
\end{equation}

\noindent{We note that Eqns. \ref{eqn1} and \ref{eqn2} written in this way do not involve the elastic scattering time except through the sheet resistance $R_\square$, which is a measured quantity and not a fitting parameter.  There are two remaining fitting parameters, $\ell_1$ and $\ell_2$.  It is very difficult to obtain reliable quantitative estimates for $\ell_\phi$, $\ell_{so}$ and $\ell_s$ from fitting a single low field MR trace to Eqn. \ref{eqn1}.  The fits are sensitive to $\ell_2$ and $\ell_1$ only when these lengths are comparable in magnitude.  In order to see this, consider the case when magnetic scattering is weak, so that we can ignore the terms involving $\ell_s$.  When $\ell_{so}>>\ell_\phi$, the fits will depend on $\ell_\phi$, but are insensitive to $\ell_{so}$.  In this case, $f(B_2,B)\sim f(B_1, B)$, and the low field MR is negative, corresponding to weak localization.  In the opposite limit, $\ell_{so}<<\ell_\phi$, the triplet term in Eqn. \ref{eqn1} is much smaller than the singlet term. Here one observes a low field MR that is positive, corresponding to weak antilocalization, but the fits again depend only on $\ell_\phi$ and are insensitive to $\ell_{so}$.  In the intermediate regime, when $\ell_{so}\sim \ell_\phi$, the MR is positive near zero field but becomes negative at higher field.  It is only in this case that the fits depend on both $\ell_\phi$ and $\ell_{so}$.  In conventional `dirty' metals such as Au, for example, one fits the low field MR in a regime where $\ell_\phi$ and $\ell_{so}$ are comparable, at higher temperatures, for example, using both lengths as fitting parameters, and then fits the MR at other temperatures with only $\ell_\phi$ as a fitting parameter, keeping $\ell_{so}$ constant to obtain $\ell_\phi$ as a function of $T$.\cite{Chandrasekhar}}

For the present LSAT/STO devices, the situation is even more complicated, as one has contributions from the classical MR, which varies with $V_g$, as well as potential magnetic scattering, so obtaining reliable quantitative estimates of $\ell_\phi$ and $\ell_{so}$ is very difficult.  Nevertheless, one can draw specific conclusions about the SOI, based on the qualitative dependence of the low field MR on $V_g$.  We first note that $\ell_\phi$, which depends on $R_\square$, is expected to decrease in magnitude as $V_g$ is decreased and $R_\square$ increases.\cite{CavigliaSOC}  At $V_g\sim100$ V then, $\ell_\phi$ is comparatively large, which agrees with our earlier observation that the system at this gate voltage is relatively cleaner, showing a large classical MR that onsets at relatively low fields, and SdH oscillations.  However, the low field MR is negative, i.e., it exhibits weak localization, which from our discussion above indicates that $\ell_{so}$ is even larger.  As $V_g$ is reduced, $\ell_\phi$ decreases, which can be inferred from the increase in the characteristic field scale, but we also observe a transition to weak antilocalization, which indicates that $\ell_{so}$ is decreasing even more rapidly with $V_g$.  Indeed, the transition between negative to positive MR near zero field occurs for $V_g\leq20$ V, the same value of $V_g$ at which we see a decline in the magnitude of the large field MR (Fig. \ref{fig2}a), which we associated earlier with the depopulation of an electron band.  This suggests that the increase in SOI is related to the depopulation of electron bands as a function of $V_g$. 



Another striking transition that occurs around the same value of $V_g$ but at millikelvin temperatures can be seen by examining the large field scale MR at various gate voltages, shown in Figs. \ref{fig2}b and \ref{fig3}b.  For $V_g\geq 60$ V, the MR is non-hysteretic; for $V_g\leq40$ V, the MR becomes progressively more hysteretic as $V_g$ is decreased.  We can distinguish two components to this hysteresis, both of which have also been observed in LAO/STO devices:\cite{Brinkman,Mehta} a large background hysteresis, which is associated with the increasing glassiness of the sample with decreasing $V_g$, and two sharp mirror-symmetric dips at low fields (Fig. \ref{fig3}c), which we associate with the presence of long-range magnetic order in the system.  Both hysteretic contributions decrease in magnitude with increasing temperature and by decreasing the sweep rate of the magnetic field (see Supplementary Information), as has also been observed in LAO/STO structures. The magnetic origin of the low field dips can be seen by applying an in-plane magnetic field, which suppresses the low field hysteretic dips, but not the high field hysteresis (Fig. \ref{fig3}d).  Thus, in concert with the depopulation of the electron band and corresponding increase in spin-orbit scattering with decreasing $V_g$, there is an onset of magnetic order in the system, strongly suggesting that this magnetic order is related to the turn-on of SOI in the system.       

To summarize, we have studied the transverse and longitudinal MR of Hall bars fabricated on a (111) LSAT/STO sample. Our Hall data suggest that holes contribute significantly to the low carrier density regime, which is also the regime in which a strong SOI develops as we reduce $V_g$. At millikelvin temperatures, the carriers in these low energy bands participate in creating ferromagnetic order, showing that SOI may play a significant role in the development of this ferromagnetic phase. We expect the SOI to be strong, since at low $V_g$ when the phase decoherence length is shortest, the SOI length is seen to be shorter still. We believe this result gives us important insight about the band structure and interactions of (111) oriented STO based systems. The tunable nature of the SOI also opens the possibility of manipulating the spin textures in the system electrically. We believe that these factors should provide an excellent opportunity for applications as well as fundamental research.

\begin{acknowledgments}
The U.S. Department of Energy, Office of Basic Energy Sciences supported the work at Northwestern University through Grant No. DE-FG02-06ER46346.  Work at NUS was supported by the MOE Tier 1 (Grants No. R-144-000-364-112 and No.R-144-000-391-114) and Singapore National Research Foundation (NRF) under the Competitive Research Programs (CRP Award No. NRF-CRP15-2015-01).  This work utilized Northwestern University Micro/Nano Fabrication Facility (NUFAB), which is supported by the State of Illinois and Northwestern University.
\end{acknowledgments}



\end{document}